\documentclass[iop]{emulateapj}

\usepackage{txfonts}
\usepackage{graphicx}
\usepackage{graphicx}
\usepackage{natbib}

\usepackage{natbib}

\shorttitle{Dynamics of chromospheric Rapid Redshifted and Blueshifted Excursions}
\shortauthors{Kuridze et al.}

\begin{document}
\title{The Dynamics of Rapid Redshifted and Blueshifted Excursions in the Solar H$\alpha$ line}
\vskip1.0truecm
\author{
D. Kuridze$^{1,4}$, V. Henriques$^{1}$, M. Mathioudakis$^{1}$, R. Erd\'elyi$^{2}$, T. V. Zaqarashvili$^{3,4}$, S. Shelyag$^{5}$, P. H. Keys$^{1,2}$, and F. P. Keenan${^1}$}
\affil
{$^1$Astrophysics Research Centre, School of Mathematics and Physics, Queen's University, Belfast, BT7~1NN, Northern Ireland, UK; e-mail: d.kuridze@qub.ac.uk} 
\affil{$^2$Solar Physics and Space Plasma Research Centre (SP$^2$RC), University of Sheffield, Hicks Building, Hounsfield Road, Sheffield S3 7RH, UK}
\affil{$^3$Space Research Institute, Austrian Academy of Sciences, Schmiedlstrasse 6, 8042 Graz, Austria}
\affil{$^4$Abastumani Astrophysical Observatory at Ilia State University, 3/5 Cholokashvili avenue, 0162 Tbilisi, Georgia}
\affil{$^5$Monash Centre for Astrophysics, School of Mathematical Sciences, Monash University, Clayton VIC 3800, Australia}
\date{received / accepted }

\begin{abstract}
We analyse high temporal and spatial resolution time-series of spectral scans of the H$\alpha$ line obtained with the
CRisp Imaging SpectroPolarimeter (CRISP) instrument mounted on the Swedish Solar Telescope.
The data reveal highly dynamic, dark, short-lived structures known as Rapid Redshifted and Blueshifted Excursions (RREs, RBEs)
that are on-disk absorption features observed in the red and blue wings of spectral lines formed in the chromosphere.
We study the dynamics of RREs and RBEs by tracking their evolution in space and time,
measuring the speed of the apparent motion, line-of-sight Doppler velocity, and transverse velocity
of individual structures.
A statistical study of their measured properties
shows that RREs and RBEs have similar occurrence rates, lifetimes, lengths, and widths.
They also display non-periodic, non-linear transverse motions perpendicular to their axes at speeds of 4 - 31 km s$^{-1}$.
Furthermore, both types of structures either appear as high speed jets and blobs
that are directed outwardly from a magnetic bright point with speeds of 50  - 150  km s$^{-1}$,
or emerge within a few seconds.
A study of the different velocity components suggests that the
transverse motions along the line-of-sight of the chromospheric flux tubes
are responsible for the formation and appearance of these redshifted/blueshifted
structures. The short lifetime and fast disappearance of the RREs/RBEs suggests that, similar to type II spicules, they are rapidly
heated to transition region or even coronal temperatures.
We speculate that the Kelvin-Helmholtz instability triggered by observed transverse motions of these structures may
be a viable mechanism for their heating.  
\end{abstract}


\section{Introduction}

The highly inhomogeneous nature of the solar chromosphere has been known since the discovery of solar spicules almost 140 years ago \citep{Sec}.
Spicules are small-scale, jet-like plasma features observed ubiquitously at the solar limb between the photosphere and the corona
\citep{rob45, beck1, beck2, ster, tsir12}. The traditional (type I) spicules have lifetimes ranging from 1 - 12 min and are characterised by rising  and falling 
motions with speeds of $\sim$20-40~km~s$^{-1}$.
Subsequent observations have revealed similar types of chromospheric fine structures such as quiet Sun mottles \citep{sue95}
and active region fibrils on the solar disk. Observations with the Solar Optical Telescope (SOT) onboard Hinode have 
identified a different type of spicule (type II), which are 
more energetic and short-lived features characterised only by upward apparent motions
\citep{dep07}.

\begin{figure*}[t]
\begin{center}
\includegraphics[width=18cm]{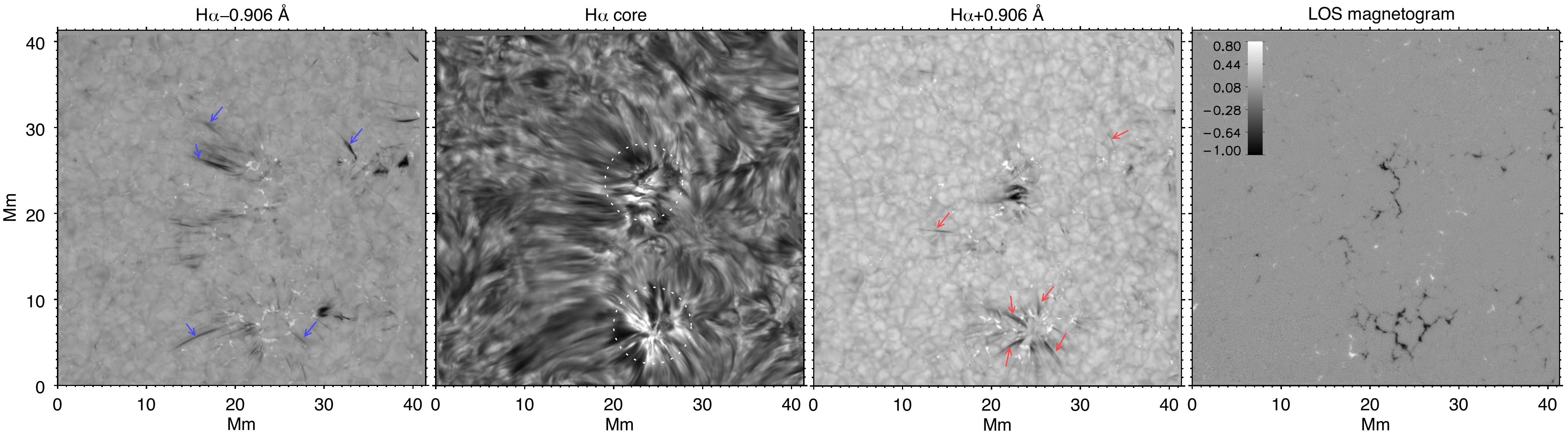}
\end{center}
\caption{H$\alpha$ core and $\pm$0.906 {\AA}  images
together with the photospheric, line-of-sight magnetogram obtained from Fe 6302 {\AA} Stokes {\it V} profiles.
Red and blue arrows indicate typical RREs and RBEs observed in the H$\alpha$ line wings.
The rosette regions where most of the RBEs/RREs are detected are highlighted with white dotted circles.
The images show that the footpoints of the RBEs/RREs correspond to photospheric bright points and strong magnetic flux concentrations.  
The colour scale in the magnetogram indicates the magnetic field strength in kilogauss. }
\label{fig1}
\end{figure*}

Another class of fine-scale chromospheric structures, called the Rapid Blueshifted Excursions (RBEs), was found
on the solar disk using high resolution, ground-based spectroscopic observations \citep{lan08}.
These are absorption features detected in the blue wings of the Ca II 8542 {\AA} and H$\alpha$ 6563 {\AA} lines \citep{lan08,voort09}.
The measured properties of RBEs are similar to those reported for type II spicules,  supporting the idea that
they are the same phenomenon viewed from different observational angles, i.e. limb vs disk.  \citep{lan08,voort09}.
These dynamic events, characterised by a blueshifted spectral line profile, occur near network boundaries mostly in areas
where H$\alpha$ 
rosettes are located. Rosettes are clusters of elongated H$\alpha$ mottles expanding radially around a common centre
over internetwork regions (Zachariadis et al. 2001; Tziotziou et al. 2003; Rouppe van der Voort et al. 2007).
However, compared to mottles, RBEs are slender  ($\sim$200 km) and short-lived ($\sim$40 s)  with higher apparent velocities ($\sim$50-150 km s$^{-1}$).
Furthermore, RBEs are characterised by rapid fading (within a few seconds) in chromospheric spectral
lines without any descending behaviour. This rapid disappearance may be the result of  fast heating to transition region and coronal temperatures  \citep{voort09}. 
Some recent observational evidence supports the heating hypothesis in limb spicules
(De Pontieu et al. 2011; Pereira et al. 2014). However, the
mechanism for this fast heating remains a mystery.  

Another important dynamical characteristic of these structures
is their transverse displacements. Like spicular structures, RBEs undergo transverse motions perpendicular to their longitudinal axis \citep{voort09}.
In relatively long-lived structures, such as type I spicules and mottles, these motions are usually observed as periodic transverse displacements, and interpreted as
magnetohydrodynamic (MHD) kink waves
\citep{zaq07, pie11, kur1, mor12, jess12}. However, it is extremely difficult to detect any such periodicity in RBEs due to their short lifetime.

\cite{sek13a} reported the presence of RBE-like absorption features in the red wing
of the Ca II 8542 {\AA} and H$\alpha$ 6563 {\AA} lines, referred to as Rapid Redshifted Excursions (RREs).
They are detected in the same regions, along chromospheric magnetic flux concentrations,
where RBEs and H$\alpha$ mottles are usually observed,
but with a lower occurrence rate. \cite{sek13a} suggested that three kinds of motion in the chromospheric flux tube, i. e. field-aligned flows,
swaying motions and torsional motions, can result in a net redshift that is manifested as RREs.
\cite{lip14}, have reported similar structures in the H$\alpha$ wing (blue- and red-wing fibrils) near the solar limb.
These structures occur with comparable occurrence rates   
without any evidence for torsional motions.
\cite{lip14} have presented arguments that at least some of the detected structures may correspond to warps in two-dimensional sheets,
which is an alternative interpretation for dynamic type II spicular-like features \citep{jud11}.

\begin{figure*}[t]
\begin{center}
\includegraphics[width=18cm]{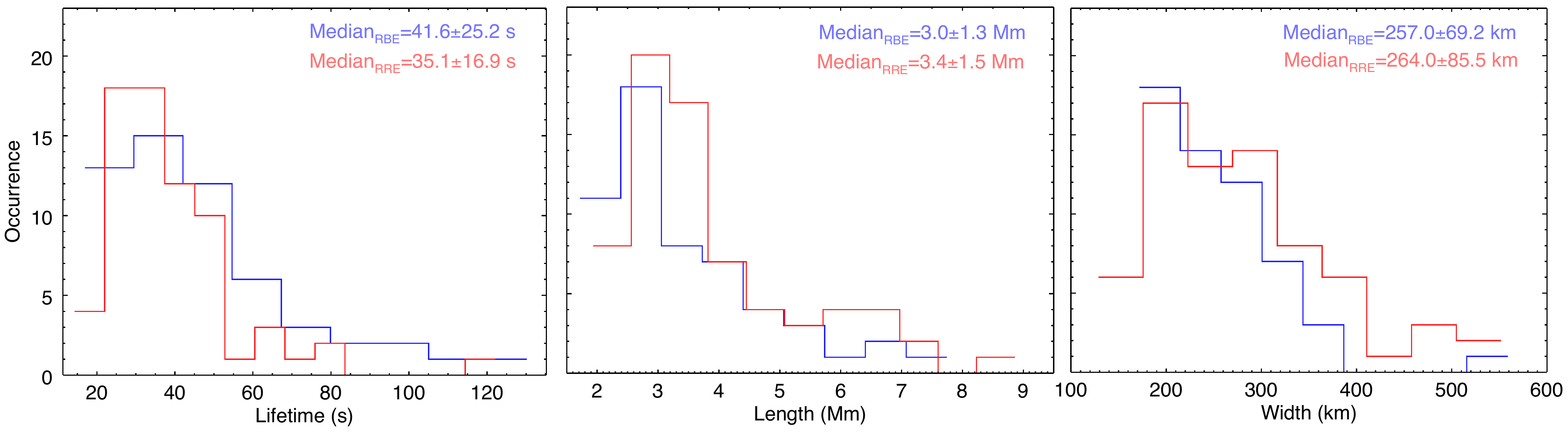}
\end{center}
\caption{The lifetime (left),  length (middle), and width (right),
for the 70 RREs (red lines) and 58 RBEs (blue lines) analysed in this study. Median values with standard deviations are also given.}
\label{fig2}
\end{figure*}

Here, we present high spatial and temporal resolution imaging spectroscopy of RREs and RBEs
in the H$\alpha$ on-disk, quiet solar chromosphere.
The structures are detected at symmetric wing positions located at  $\pm$0.906 {\AA} from the H$\alpha$ line core.
Some of them are high-speed, dark, blob-like features moving upwardly along magnetic flux tubes.
To investigate their dynamical characteristics and spatio-temporal evolution we perform detail measurements of their 
line-of-sight (LOS) velocities, transverse velocities in the image
plane and apparent velocities of the individual structures. Using the measured properties, 
the interrelations between RREs and RBEs and the formation of their spectral line asymmetries are investigated.
We discuss which mechanism could be responsible for the rapid heating, and hence the observed
fast disappearance of these structures, as well as show the presence of similar features in numerical MHD simulations of the solar atmosphere.

\section{Observations and data reduction}
\label{sect:setup}

Observations were undertaken between 09:06 and 09:35 UT on 2013 May 5 at disk centre, on the edge of a coronal hole, with the CRisp Imaging SpectroPolarimeter \citep[CRISP;][]{shr06,shr08} instrument, 
mounted on the 1-m Swedish Solar Telescope \citep[SST;][]{shr03}.
The dataset includes spectral imaging in the H$\alpha$~6563~{\AA} and Fe~6302~{\AA} lines. Adaptive optics were used throughout the observations consisting of a tip-tilt mirror and a 85-electrode deformable mirror setup that is an upgrade of the system described in \cite{shr03}.
This system provided a near constant ``lock" and all reconstructed frames benefited from the adaptive-optics correction.

All data were reconstructed with Multi-Object Multi-Frame Blind Deconvolution \citep[MOMFBD;][]{noo05,lof11}, using 51~Karhunen-Lo\`{e}ve modes sorted by order of atmospheric significance and using $88\times88$ pixel subfields in the red. 
A prototype of the code published by \cite{rod14} was used before and after MOMFBD. Reflectivity fitting, due to the broadness of H-alpha, was not used as justified in \cite{rod13}. 
The images, reconstructed from the narrowband wavelengths, were aligned at the smallest scales by using the method described by \cite{hen13}.
This employs cross-correlation between auxiliary wide-band channels, obtained from an extended MOMFBD scheme, to account for different residual small-scale seeing distortions and different reconstruction PSFs.
Data were composed into time-series having been de-rotated, aligned, and finally destretched as in \cite{shi94}. 

Spatial sampling is 0$''$.0592 pixel$^{-1}$  and the spatial resolution up to 0$''$.16 in H$\alpha$ over the field-of-view (FOV) of 41$\times$41 Mm.
The H$\alpha$ line scan consist of 7 positions (-0.906, -0.543, -0.362, 0.000, 0.362, 0.543, +0.906 {\AA} from line core) corresponding to a velocity range of -41 to +41 km s$^{-1}$ in velocity.
Following reconstruction, the cadence of a full spectral scan was 1.34 s.
CRISP also undertook photospheric Fe~6301 and 6302~{\AA} spectral scans with 16 wavelength positions per line every  $\sim$5 minutes over the same FOV to obtain full Stokes profiles.
The Stokes {\it V} component was used to construct LOS magnetograms employing the centre of gravity method \citep{ree79,uit03}.

The widget-based CRIsp SPectral EXplorer \citep[CRISPEX;][]{vis12} and Timeslice ANAlysis tools (TANAT) were used to perform the data analysis and the measurements presented in this study.


\section{Analysis and Results}

Figure~\ref{fig1} shows images in the H$\alpha$ line core and $\pm$0.906 {\AA}  
from line core  
together with the photospheric LOS magnetogram obtained from the Fe 6302 {\AA} Stokes {\it V} profiles.
The H$\alpha$ line core image contains two large rosette structures, highlighted with white dotted circles (second panel of Figure~\ref{fig1}), 
where most of the chromospheric flux tubes, and hence the fine-scale structures, are concentrated.
The roots of the rosette structures are co-spatial with strong magnetic field concentrations that
outline the intergranular lanes.  
The LOS magnetogram shows that the FOV consists mostly of unipolar patches with magnetic field strength in the kilogauss range.


We selected dark absorption features manually (i.e by visual inspection) in the far 
red and blue wings at H$\alpha\pm$0.906~{\AA}
that are longer than 1.5 Mm for at least 10 seconds, and have an aspect ratio (length/width) greater than $\sim$5 at maximum extension.
We have excluded granular boundaries and so-called swirls, characterised by vortex-like behaviour, from our selection.  
Based on these criteria, we detected 70 features in the far red wing at H$\alpha+$0.906~{\AA} and 58 features in the far blue wing  at H$\alpha-$0.906~{\AA}, identified as RREs and RBEs respectively.
Examples of these dark absorbing structures, originating from magnetic bright points along chromospheric flux tubes,
are indicated by the red and blue arrows in the H$\alpha$ line wing images (Figure~\ref{fig1}).
It should be noted that the number of RBEs/RREs detected in this work is much smaller than the number of detections by an automated algorithm used by other authors \citep[see e.g.,][]{sek12}.
However, our selection is restricted to features that are
clearly distinguishable, fully observable from edge to edge, and fully traceable in time in the far wing images of the H$\alpha$ time sequence. 
This reduces the number of detections but gives us a more robust dataset to investigate the different velocity components and the spatial-temporal evolution which is the ultimate goal of this work.
We also want to note that the total number of detected RREs is higher than the number of RBEs (70 vs 58), we believe that RREs and RBEs appear here with a similar occurrence rate.
The difference in numbers may be attributed to the 
lower contrast between the background and RBEs.

We have measured each structure's lifetime as the time interval between its first appearance and complete disappearance; the
maximum projected length as the maximum visible extension with manually identified lower and upper ends of the structure during its lifetime; 
the width as the visible diameter of the central part of  the detected structure at the time of maximum extension.
The width measured with this method agrees very well with full-width at half-maximum of Gaussian-fit trials performed on the cross-section of the flux profiles. 

The results show that RREs and RBEs have almost identical lifetimes, widths and lengths (Figure~\ref{fig2}).
Despite these similarities, the analysis does not show that the appearance or disappearance of RREs is correlated with the appearance or disappearance of RBEs suggesting that they are independent features.
Similarly, there is no one-to-one relation between them and H$\alpha$ line core mottles.

\begin{figure*}[t]
\begin{center}
\includegraphics[width=18cm]{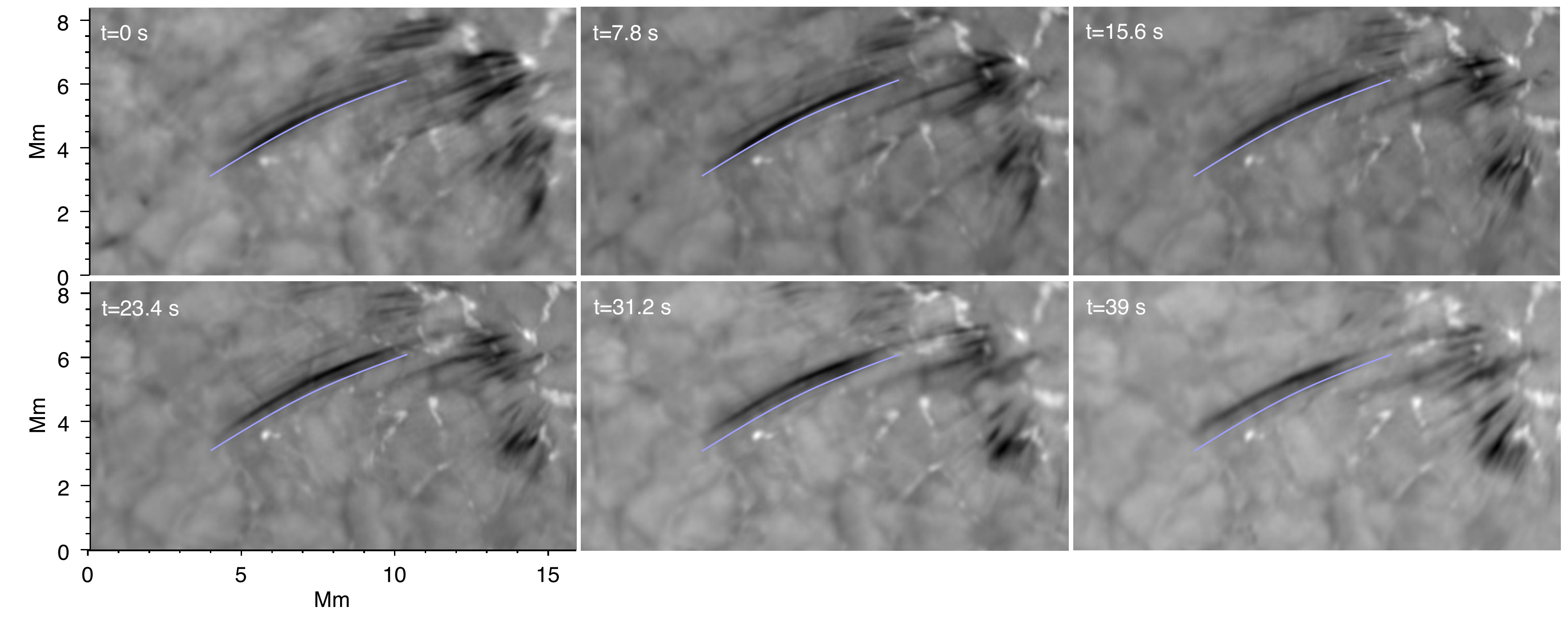}
\end{center}
\caption{
A typical example of the non-periodic transverse displacement of an RBE with a displacement of $\sim$500~km and a transverse velocity of $\sim$13~km~s$^{-1}$. Blue lines show the initial position of the structure for reference.}
\label{fig3}
\end{figure*}

The detected structures either appear as upward-directed high speed jets and blobs, or emerge suddenly within a few time resolution elements ($\sim$1-5 s).
Many of them (13 RREs and 22 RBEs) display transverse motions perpendicular to their axis in the image plane.
At the end of their lifetime the structures disappear extremely quickly
without any apparent downward motion.
We note that RREs/RBEs detected in the $\pm$0.906{\AA} images, are normally also visible in neighbouring $\pm$0.543{\AA} images. They appear and disappear simultaneously in both wavelength positions.
Transverse velocities and transverse displacements
are measured using time-distance (t-d) diagrams generated along the cuts perpendicular to the structure's axes.
Figure~\ref{fig3} illustrates a typical example of transverse displacement of a selected RBE.
We~ do~ not ~ see  ~periodic  behaviour for  the  ~detected ~ transverse ~ motions~ (i. e., there are no swings from side to side).
However, it should be noted that the structures disappear in the H$\alpha$ images before the transverse motions come to a halt. 
Transverse velocities are in the range of 4 - 22~km~s$^{-1}$, with displacements between 250 and 1200 km (Figure~\ref{fig4}).
It is clear that the transverse displacements are greater than the width of the structures ($\sim$250~km), implying that
the motions are strongly nonlinear.

The apparent velocities of the RREs and RBEs projected on the image plane are determined using t-d plots, generated along the path of these motions.
Velocity estimates are in the range 50 - 150 km~s$^{-1}$ (Figure~\ref{fig4}) which are Alfv\'enic and super-Alfv\'enic in the chromosphere. 
The apparent absolute velocities of RREs and RBEs are in a similar range, although the number of structures where reliable measurements 
were possible are different (36 vs 5 for RREs and RBEs, respectively).
We believe that the higher contrast of the RREs makes the
apparent motions in them more easily measurable than in RBEs.
The H$\alpha$ data presented here shows no evidence for any downward motions either in the RBEs or in the RREs.

Some of the detected RREs/RBEs appear as high-speed blobs moving upwardly along the same path where other chromospheric
fine structures are detected.  
Figure~\ref{fig5} shows the temporal evolution of a typical dark blob detected in the red wing at H$\alpha$+0.906~{\AA}.
The blob starts from the bright point and moves upward along a curved trajectory
with an apparent propagation speed of $\sim$110 km~s$^{-1}$.

The large red and blue line profile asymmetry, which is the main characteristic of RREs and RBEs, means that they have a LOS velocity component. 
Values of LOS velocities for each individual feature are calculated using Doppler signals provided by their H$\alpha$ spectral profile.
For these measurements we use the first moment with respect to wavelength method described by \cite{voort09}. 
A box is defined for each manually detected RRE/RBE that covers the whole structure when it has maximum extension of $\pm$5 pixels, and calculate the $v_{LOS}(x,y,i)$,
where $x$ and $y$ are transverse and longitudinal coordinates of this box, and $i=j,...,k$ so that $j$ and $k$ are the number of frames
where the structure first appears and disappears completely, respectively. 
Doppler velocities were compensated after computation for etalon cavity errors.
Our analysis shows that the LOS velocities are approximately constant  along the lengths of RREs/RBEs.
Furthermore, they are almost regular and do not show any type of periodic behaviour over their lifetime. 
 An absolute value of these velocities, averaged along the structures lengths, are in the range of 21-34 km~s$^{-1}$.
The distribution of LOS velocities is shown in Figure~\ref{fig4} (bottom right panel) where again both type of structures have similar LOS velocities.

We inspected the locations where RBEs or RREs were detected a few minutes before and after the appearance/disappearance of each structure and looked for evidence in both the LOS Doppler velocity signal and intensity in the line wings. We found no significant signal and therefore no evidence of returning or periodic motions.

\section{Discussion}

\subsection{Transverse displacements}

A wide range of transverse motions have been studied thoroughly in chromospheric fine structures such as type
I and II spicules \citep{zaq07, zaq09, oka11}, mottles  \citep{kur1, mor12,kur13}, fibrils \citep{pie11}, and RBEs \citep{voort09,sek13a}. 
These motions frequently reveal periodic behaviour which is attributed to MHD kink waves \citep{spr82}.
Reported wave periods in the relatively long-lived chromospheric features, such as type I spicules, mottles and fibrils, are in the range 1 - 5 minutes.
The mean period of the kink waves in the H$\alpha$ line core mottles are $\sim$3~min \citep{kur1, mor12}.
RREs/RBEs have similar geometries and spatial scales (length and width) as the H$\alpha$ mottles.
Furthermore, as they occur at the same place and their footprints are almost co-spatial,
it may be expected that they will have the same wave drivers. Thus waves with similar periods may be expected to be excited in them as well.
However, RREs/RBEs are short-lived, compared to mottles, with mean lifetimes much lower
than the typical periods of transverse kink waves in chromospheric fine structures.
This makes it extremely challenging to determine the periodicity of these motions in RREs and RBEs.  
We note that some evidence of periodic transverse motions in RBEs has been reported by \cite{sek13b}.
\begin{figure*}[t]
\begin{center}
\includegraphics[width=15cm]{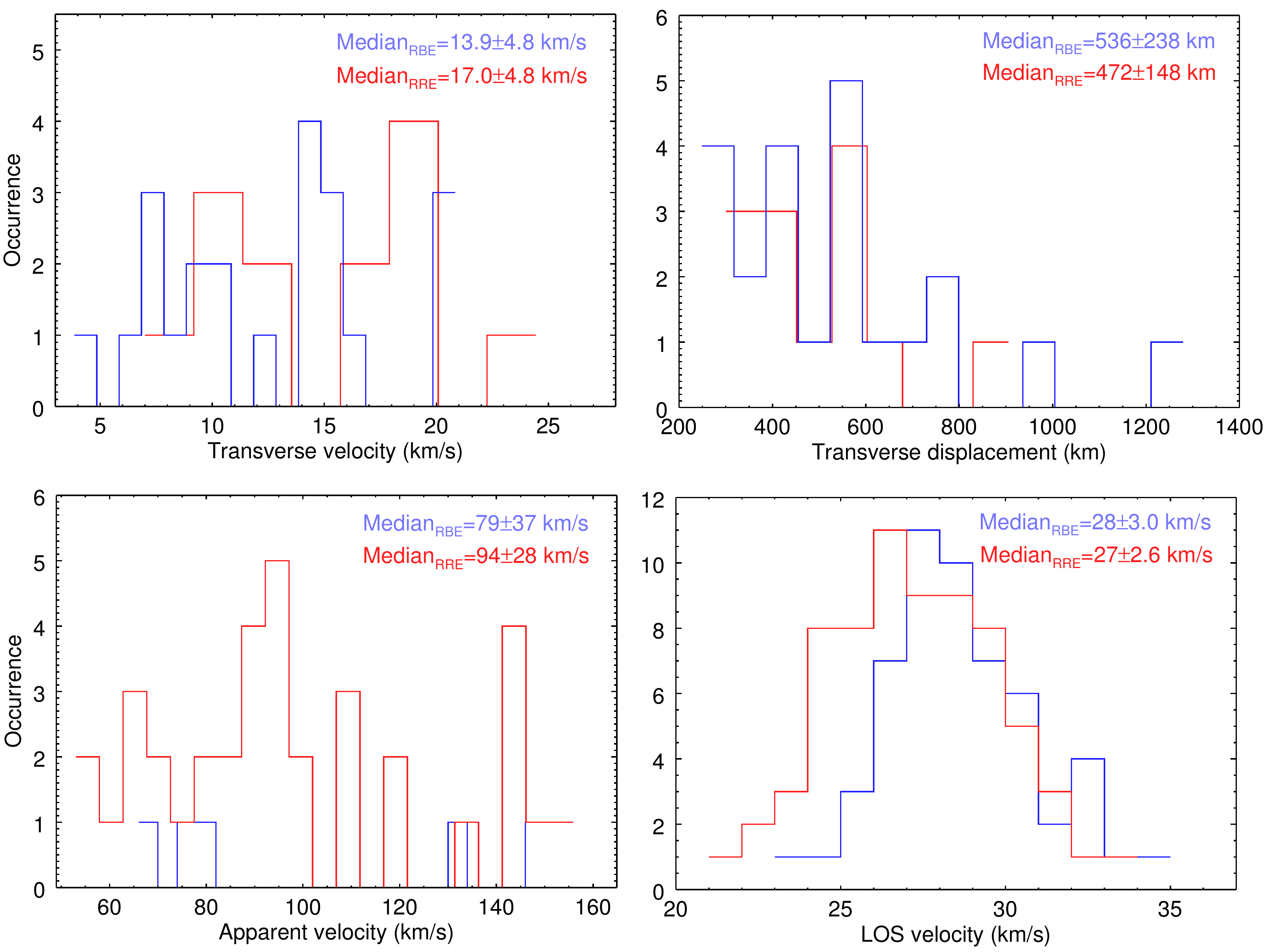}
\end{center}
\caption{
Distribution of the average LOS velocity (absolute value), apparent velocity, transverse velocity (in the image plane)
and transverse displacement amplitude for the analysed RREs, RBEs (red and blue lines, respectively). Median values with standard deviations are also given.}
\label{fig4}
\end{figure*}
The transverse motions analysed in this work do not show full swings (i.e., motions from left to right and back again) that would be the manifestation of kink waves. 
RREs and RBEs display a nonlinear transverse motion in only one direction that does not come to a halt before the structure disappears.
This suggests 
that the reason for the lack of a full wave cycle is due to the relatively short lifetimes of the RREs and RBEs. 
It should be noted that the strong damping of these motions could be an alternative explanation for their non-periodic nature. 
However, scaling laws of existing damping models, such as e. g. resonant absorption, phase mixing, wave leakage or mode conversion,
indicate that the damping timescales should be longer than the wave periods \citep{goo02, ofm02, goo14}. This suggests that a periodicity could still have been seen.
It could be argued that the observed transverse motions are not waves, but just transverse displacement of the flux tube as a whole from its initial position.
\cite{has99} have shown that when transverse motions are excited by a photospheric granule, and
the interaction timescales between the granule and the flux tube is comparable or greater than the chromospheric cutoff period,
there is no kink wave. Instead, the whole tube is simply displaced from its initial location at all heights.
Unfortunately, the dataset analysed in this work does not allow us to determine the excitation mechanisms,
and hence the exact nature of the detected transverse motions.

\subsection{Apparent and LOS Velocities}

The observations presented here show that
RREs and RBEs either display apparent motions
that are directed outwardly from the magnetic bright point,
or emerge suddenly within a few seconds.  
Our very high cadence (1.3~s) allowed us to estimate apparent motions for some of the 
RREs and RBEs with speeds ranging between 50 - 150 km~s$^{-1}$ (Figure~\ref{fig4}).
Similar behaviour in RREs and type II spicules has been reported by other authors \citep{voort09,sek13a}.
The Alfv\'enic/super-Alfv\'enic apparent velocities measured here, and the sudden appearance of the structures, are their most puzzling aspects. 
One possible explanation is that they are formed as a result of upward directed plasma flows along chromospheric flux tubes.  
A number of recent numerical simulations have shown that the high-speed upflow patterns in the chromosphere can be generated
by photospheric vortex motions in magnetic features and Lorentz forces \citep{good12, kit13}. However, the extremely fast appearance
(with speeds of more than $\sim$700~km~s$^{-1}$) is still unexplained by this flow scenario. Furthermore, it does not explain the appearance of
large numbers of redshifted features (RREs) at the solar disk centre. The sources of the Doppler signals of the chromospheric flux tube could be
field-aligned flows  and/or the transverse motions along the LOS.  It must be noted that torsional motions can also contribute to the line shift,
but we have not found any evidence of torsional motions in our dataset.
Field-aligned flows that appear from the bright points and move outwardly along a magnetic flux tube can produce blueshifted Doppler signals, $v_{dopp}\rightarrow blue$,
if the tilt angle ($\theta$) with respect to the observer is $<$90$^{\circ}$, $v_{dopp}\rightarrow0$ if $\theta\approx$90$^{\circ}$, and $v_{dopp}\rightarrow red$ if $\theta>$90$^{\circ}$.
As the observed quiet Sun region is almost unipolar  (right panel of Figure~\ref{fig1})
it is expected that the region is dominated by open flux tubes which suggests that they do not have $\theta<$90$^{\circ}$ at the disk centre.
Thus, flows can only produce blueshifted signals whereas transverse motions can generate both red and blue Doppler-shifts without any preference.
Hence, if the apparent motions are plasma flows, then there are more sources a blueshifted Doppler signals.
There should therefore be a prevalence of RBEs and the absolute values of the Doppler velocities for the RBEs should be higher than 
of RREs. However, we could find no evidence that would substantiate this scenario.  
By contrast, the number of detected RREs is higher than that of RBEs and they have similar absolute values of LOS velocities
(bottom right panel of Figure~\ref{fig4}). This indicates that field aligned flows is an unlikely explanation for the fast appearance of the RREs and RBEs.

 \begin{figure*}[t]
\begin{center}
\includegraphics[width=18cm]{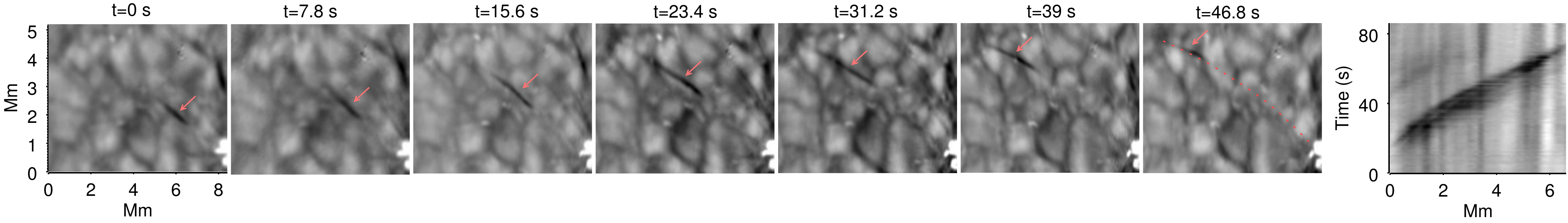}
\end{center}
\caption{Time-sequence of a typical dark blob detected in the red wing at H$\alpha$+0.906~{\AA} (42 km~s$^{-1}$).
The blob starts from the network brightening and moves upward along the curved trajectory highlighted with the red dotted line.
The corresponding t-d diagram (right panel) estimates
an apparent propagation speed of $\sim$110 km~s$^{-1}$.}
\label{fig5}
\end{figure*}

An alternative interpretation for the high apparent velocities and rapid, sudden appearance of these features is based on a wave scenario. 
The basic idea is that upward propagating kink waves/disturbances along the flux tube can be seen as Doppler signals in the form
of RREs/RBEs appearing with Alfv\'enic speed. Indeed, the observed high-speed
redshifted/blueshifted jet- and blob-like features may be interpreted in terms of
propagating transverse pulses/disturbances travelling with a phase speed which could be close to the chromospheric Alfv\'en speed. On the other hand,
superposition of the opposite-directed kink waves may result in a wave with a very high/infinite phase speed 
(standing waves) that can be manifested as the sudden appearance of RREs/RBEs. 
Both propagating (upward and downward) and standing, transverse oscillations have been reported
in type II spicules at the limb \citep{oka11} and mottles on the disk \citep{kur13}. However, there is no evidence of downward propagating signals
in RBEs and RREs. Furthermore, as noted by \cite{lip14}, standing waves require a finite time to set up. 
This time is estimated as $t\sim l/v_{A}$~s \citep{jud12}, where $l$ and $v_{A}$ is the structure's length and chromospheric Alfv\'en speed, respectively.
RREs and RBEs have typical lengths of $\sim$3~Mm, and $v_{A}$ is estimated as $\sim$20-80 km~s$^{-1}$ in the chromosphere.  
Hence $t$ is $\sim$40 - 150 s which is comparable/greater than their typical lifetime ($\sim$40~s).
It is therefore doubtful that  standing wave patterns, generated by the superposition of oppositely directed waves, are responsible for the
observed fast appearance of the RREs/RBEs. 

The main aspect of the reported transverse motions in RREs and RBEs is that the structure is displaced as a whole,
i. e. every single part of their visible length is moved in phase, in the same direction and with the same speed.
These motions ($v_{{tr}_{\bot}}$) are detected in the image plane. However,
it is expected that the motions should also have a LOS component ($v_{tr_{LOS}}$)
and some should also be polarised in the LOS plane.
If the flux tubes start to move along the LOS in a similar way to what is detected in the image plane (Figure~\ref{fig3}),
then they should appear suddenly in the H$\alpha$ line wings as redshifted or blueshifted
features depending on the direction of the LOS velocity with respect to the observer.
We propose that the observed RREs and RBEs are manifestations of the chromospheric flux tubes transverse motions along the LOS.  Similar to the transverse motions in the image plane,
LOS Doppler signals and thus the associated transverse motions
do not show any evidence of periodic behaviour,
e.g. periodic variation or the change of the flux tubes Doppler-shift sign.
It must be noted that the obtained median values of $v_{tr_{LOS}}$ ($\sim$27-28 km~s$^{-1}$ for RREs and RBEs,
respectively) are higher compared to the median of $v_{{tr}_{\bot}}$  ($\sim$14-17 km~s$^{-1}$ for RREs and RBEs, respectively).
However,
as the analysed structures are detected at the far red and blue wing positions
of H$\alpha$ ($\pm$0.906~{\AA} corresponding to $\pm$41~km~s$^{-1}$),
$v_{tr_{LOS}}$  derived from
spectroscopic observations are biased toward detecting higher velocity events,
which may be the reason for the discrepancy between the average values of $v_{tr_{LOS}}$ and $v_{tr_{\bot}}$.

\subsection{Stability}

Transverse motions of the RREs/RBEs can create velocity discontinuities between
the surface of the flux tube and surrounding media, which may trigger the Kelvin-Helmholtz Instability (KHI).
The component of the magnetic field parallel to the discontinuity tends to stabilize sub-Alfv\'enic flows \citep{chan},
while the perpendicular component has no effect on KHI \citep{sen}. This suggests that the tube moving with
any speed at an angle to the magnetic field is unstable to KHI for sufficiently high azimuthal wave number $m$ \citep{zaq14}.
The observed velocities of the LOS transverse motions of the RREs/RBEs ($\sim$ 21-34 km~s$^{-1}$)
could be higher than the local Alfv\'en speed, which may significantly enhance the growth rate of KHI.

The growth rate of KHI of a transversally moving magnetic structure can be easily estimated in the case of slab geometry. The dispersion relation of antisymmetric incompressible perturbations for a magnetic slab of half-width $d$ moving transversally in its plane can be written as \citep{zaq11}
$$
\rho_{0}\left[\left(\textbf{\ensuremath{\omega}}-\textbf{k\ensuremath{\cdot}U}\right)^{2}-\left(\textbf{k\ensuremath{\cdot}V}_{A0}\right)^{2}\right]\coth(kd)+
$$
\begin{equation}
~~~~~~~~~~~~~~~+\rho_{e}\left[\omega^{2}-\left(\textbf{k\ensuremath{\cdot}V}_{Ae}\right)^{2}\right]=0
\end{equation}
where $\rho_0$, $\rho_e$, and ${\bf V_{A0}}$, ${\bf V_{Ae}}$ are the densities and Alfv\'en speeds inside and outside the slab, respectively,
${\bf U}$ is the unperturbed velocity (transverse velocity of flux tube in our case) and ${\bf k}$ is the wave vector. For symmetric perturbations $\coth(kd)$ should be replaced by $\tanh(kd)$. For $\left |kd\right |>1$, both $\coth(kd)$ and $\tanh(kd)$ are constant and equal to unity. Then Eq. (1) is simplified and using $\textbf{k\ensuremath{\cdot}V}_{A0}=\textbf{k\ensuremath{\cdot}V}_{Ae}\approx 0$ we get the imaginary part of perturbations, which is actually the growth rate, as $\ensuremath{\omega_i}\approx kU\sqrt{\rho_e/\rho_0}$. Thus the growth time of KHI depends on the transverse velocity, the density contrast and the spatial scale of perturbations. A transverse velocity of 20 km s$^{-1}$, a spatial scale of the order of tube width of 200 km and the density contrast of $\rho_e/\rho_0=0.1$ leads to the growth time of 5 s.

The generated KH vortices may lead to enhanced MHD turbulence near the tube boundaries through nonlinear cascade.
Shear flow energies then can be rapidly transferred to small scales, where it will dissipate and heat the surrounding plasma. This explains the fast heating of the chromospheric flux tubes
and hence observed sudden disappearance of RREs/RBEs. However, this is a speculative interpretation and detailed numerical simulations are required to assess the importance of KHI on the stability of these features.

\begin{figure*}
\begin{center}
\includegraphics[width=18cm]{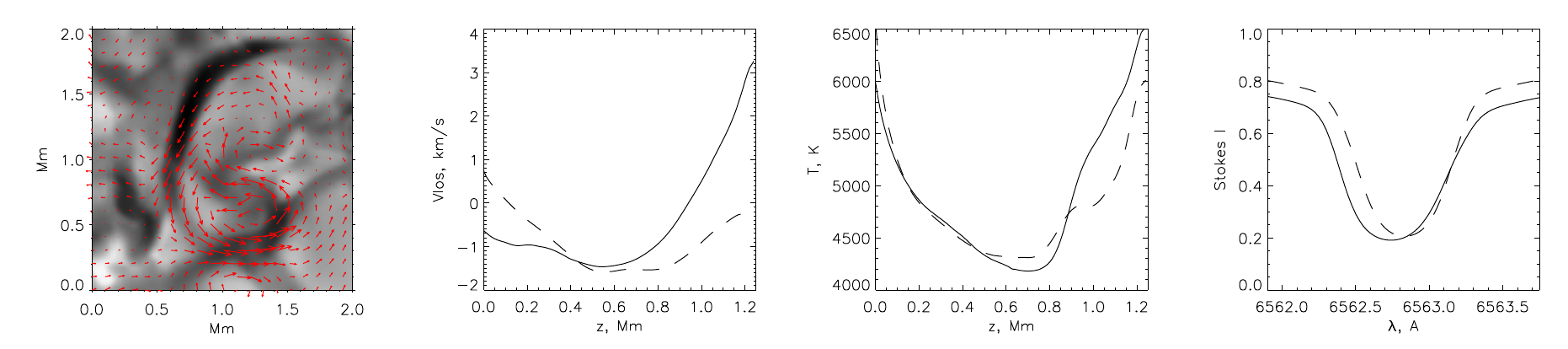}
\end{center}
\caption{ Example of a simulated RBE. First panel: Intensity in the H$\alpha$ blue wing. Second panel: line-of-sight velocity for the average RBE (solid curve) compared to regions not showing the RBE (dashed curve). Third panel: temperature profiles for the same averages. Fourth panel: average H$\alpha$ Stokes-$I$ profiles for RBE (solid curve) and non-RBE (dashed curve).
The averaging was carried out for 50 random RBE and non-RBE locations in the simulation domain. The horizontal flow structure around the simulated RBE is shown with arrows in the first panel.}
\label{fig6}
\end{figure*}


\subsection{Numerical modelling}


We carried out numerical modelling using the radiative MHD code MuRAM \citep{voegler1}. 
In a numerical domain, which covers the top $2~\mathrm{Mm}$ of the convection zone and the bottom 
$1.2~\mathrm{Mm}$ of the solar atmosphere, transparent bottom and semi-transparent (allowing outflows) top boundary conditions were used. 
A non-magnetic convection snapshot was employed as an initial configuration, 
where $200~\mathrm{G}$ uniform unipolar vertical magnetic field was injected. 
The numerical domain was then let to develop for about 20 minutes of physical time. 
One of the developed snapshots of the simulation was used to carry out H$\alpha$ diagnostics with the NICOLE code in non-LTE \citep{socas1}. 
The diagnostics showed elongated features similar to RBEs with lifetimes of 
$\sim40~\mathrm{s}$ and a length of few Mm, naturally appearing in the simulations (see Figure~\ref{fig6}, first panel).
These features are produced by strong, short-lived localised plasma
motions upward along the field lines of magnetic flux concentrations in the low-$\beta$ plasma (Figure~\ref{fig6}, second panel).
The RBEs observed in the simulations do not appear aligned with the magnetic field, however they lie at constant Alfv\'enic surfaces
and are possible signatures of dissipation of torsional Alfv\'en waves
\citep[][also Figure~\ref{fig6}, third panel]{shelyag2013, shelyag2014}. Despite some qualitative similarities between
the observations and the simulated H$\alpha$ line profiles (see Figure~\ref{fig6}, fourth panel) as well as their temporal
and spatial characteristics, it is still not fully clear if the features seen in the simulations are the observed RBEs.
There is a significant difference between the shape of simulated and observed RBEs, which may be a result of forcing the
magnetic field to be vertical at the top of the simulation domain. However, if it is assumed that the observed RBE structures
track the field lines in strongly inclined (relative to the LOS) magnetic field concentrations, the torsional velocity component would cause the red- and blueshift of the H$\alpha$ profile.
A torsional velocity component is present in the flux tubes in the simulations (Figure~\ref{fig6}, first panel) and is linked to Alfv\'en waves propagating from the base of the photosphere \citep{shelyag2013}.
Furthermore, it should be noted that no strong RREs are observed in the model due to the top boundary condition in the MHD simulation,
which does not permit downflows across the boundary, while the outflows (and, correspondingly, RBEs) are allowed.


\section{Concluding remarks}

We present a detailed study
of the temporal evolution of RREs and RBEs and their velocity components. The latter suggest that the transverse
motions along the line-of-sight are responsible for their formation and sudden
appearance in the H$\alpha$ line wings. RREs and RBEs that
appear in the form of jets and blobs with Alfv\'enic velocities may be interpreted as transverse pulses/disturbances propagating along magnetic flux tubes.
We propose that the transverse motions of chromospheric flux tubes can develop
the KHI at the tube boundaries, which may lead to the rapid heating 
and sudden disappearance of RREs/RBEs.
Numerical MHD modelling shows features qualitatively similar to the observed RBE. However,
further study is necessary to provide unambiguous link between the simulated and observed events. A more detailed study
of the simulated RBEs at different solar disk positions and angles between the magnetic field direction and the LOS is required.

\begin{acknowledgements}
The research leading to these results has received funding from the European Community's Seventh Framework Programme (FP7/2007-2013) under grant agreement no. 606862 (F-CHROMA). This research was supported by the SOLARNET project (www.solarnet-east.eu), funded by the European Commissions FP7 Capacities Program under the Grant
Agreement 312495. The work of T.Z. was supported by FP7-PEOPLE-2010-IRSES-269299 project- SOLSPANET, by Shota Rustaveli National Science
Foundation grant DI/14/6-310/12 and by the Austrian Fonds zur Förderung der
wissenschaftlichen Forschung under project P26181-N27.
Dr Shelyag gratefully thanks Centre for Astrophysics \& Supercomputing of Swinburne University of Technology (Australia), NCI National Facility systems at the Australian
National University, supported by Astronomy Australia Limited, and the Multi-modal Australian ScienceS
Imaging and Visualisation Environment (MASSIVE) for the computational resources provided.
Dr Shelyag is the recipient of an Australian Research Councils Future Fellowship (project number FT120100057).
\end{acknowledgements}

\end{document}